\documentclass{article}

\PassOptionsToPackage{numbers, compress}{natbib}

\usepackage[preprint]{neurips_2026}
\usepackage[utf8]{inputenc} % allow utf-8 input
\usepackage[T1]{fontenc}    % use 8-bit T1 fonts
\usepackage{xurl}
\usepackage{hyperref}       % hyperlinks
\usepackage{url}            % simple URL typesetting
\usepackage{booktabs}       % professional-quality tables
\usepackage{amsfonts}       % blackboard math symbols
\usepackage{nicefrac}       % compact symbols for 1/2, etc.
\usepackage{microtype}      % microtypography
\usepackage{xcolor}         % colors
\title{Open Problems in AI Risk Modeling: \\
Insights from a Workshop on the Technical Foundations of AI Risk Modeling}

\author{%
  Krystal Jackson\dag \thanks{This work completed primarily during affiliation with the Center for Long-Term Cybersecurity} \\
  Institute for Security and Technology \\
  \And
  Deepika Raman\dag \\
  Center for Long-Term Cybersecurity \\
  \And
  Jakub Kryś \\
  SaferAI \\
  \And
  Sean P.~Fillingham \\
  Independent Researcher \\ 
  \And
  Jack Kengott \\
  SaferAI \\
  \And
  Andrew Lohn \\
  Center for Security and Emerging Technology \\
  \And
  Nada Madkour \\
  Center for Long-Term Cybersecurity \\
  \And
  Henry Papadatos \\
  SaferAI \\
  \And
  James Sykes \\
  University of Warwick \\
  \And
  Anna Katariina Wisakanto \\
  Center for AI Risk Management \& Alignment  \\
  \And
  Malcolm Murray\dag \\
  SaferAI \\
}

\begin{document}

\maketitle

\begingroup
\renewcommand{\thefootnote}{\dag}
\footnotetext{Corresponding authors: krystal@securityandtechnology.org, deepika.raman@berkeley.edu, malcolm@safer-ai.org}
\endgroup

\begin{abstract}
  We investigate the design of robust risk models to assess societal risks posed by advanced AI systems, an emerging area in AI governance. Many regulatory proposals increasingly require systemic risk assessment, but in the absence of rigorous quantitative methods, the question remains what state of the art risk modeling should look like in practice. We identify the key methodological and institutional challenges that currently limit the adoption of risk modeling. We review five research traditions that inform this problem: probabilistic risk assessment, catastrophic AI risk analysis, cybersecurity risk quantification, Bayesian causal inference, and threshold-based governance. We compare two leading proposals, scenario-based risk estimation and Bayesian network-based threshold setting. Drawing on a workshop with 22 experts and subsequent analysis, we identify a structured agenda of open questions concerning model structure, scope, evidence integration, validation, and governance. We close by outlining priorities for progress, arguing that it will depend on integrating quantitative modeling with independent evaluation, transparent and tiered disclosure, and institutions capable of maintaining and updating risk models over time.
\end{abstract}

%%%%%%%%%%%Table of contents %%%%%%
\newpage
\tableofcontents
\newpage

%%%%%%%%%%%% Introduction %%%%%%%%%
\section{Introduction}
\label{sec:introduction}
The need for rigorous and quantitative methods to assess the societal risks posed by AI systems has become a pressing challenge for AI governance. Without standardized risk assessment methodologies, regulators are left with qualitative descriptions and a variety of competing approaches. Regulatory frameworks such as the EU AI Act, California SB 53, and the Seoul AI Safety Summit's Frontier AI Safety Commitments \citep{eu_gpai_code2026,ca_sb53_2025,seoul_ministerial_2024} have all recommended that frontier AI developers conduct systemic risk assessments, but largely leave the methodological details of those assessments undefined. The EU AI Act's Code of Practice specifically calls for signatories to conduct both systemic risk estimation and systemic risk modeling. The Code calls for using “state of the art” methods, but does not define what these are.\footnote{Stated in Measure 3.3 Systemic Risk Modeling. Applicable to providers of the most advanced models that are subject to the AI Act's obligations for providers of general-purpose AI models with systemic risk under Article 55 of the AI Act.} Currently, there is no consensus on what the state of the art is \citep{eder2026_sota_gpai} or what best practices for AI risk modeling should involve, in part because there are still a variety of technical and governance challenges around the operationalization of these methods. This paper aims to summarize the existing literature on AI risk modeling and highlight current gaps in research and practice that, once addressed, would allow robust risk modeling methods and tools to be developed.

Other safety-critical industries have well-established probabilistic risk assessment methods. However, these approaches have limited utility when applied to frontier AI systems. In many of the safety-critical domains where risk modeling is employed, such as nuclear and aerospace, failure modes can be clearly specified and bounded. Crucially, this condition does not hold for frontier AI systems, which have general-purpose capabilities and operate in open-ended environments. The MIT AI Risk Repository \citep{slattery2024_airisk_arxiv} catalogs a large number of diverse potential and observed risks posed by AI systems, and includes many overlapping entries. This breadth illustrates the scale and heterogeneity of the AI risk landscape and the extent to which AI risks are less standardized than those in other domains. Current risk frameworks and modeling methods cannot accommodate the number of variables and dynamic interactions between them required to capture this risk space, let alone adaptively adjust the probability distributions from which their estimates are derived. Furthermore, there are important limitations to how risk models can be used that depend on the final design decisions made and require the creation of new methods of governance. As a result, new research efforts have emerged with the goal of enabling risk modeling to meet these challenges. The methodological frontier we describe in this paper is a convergence of at least five distinct research traditions: probabilistic safety engineering, catastrophic AI risk analysis, cybersecurity risk quantification, Bayesian causal inference, and the study of intolerable risk thresholds and red lines. No single tradition is sufficient on its own, and the hybrid and integrative approach required is still being constructed.

This paper is organized around two objectives.

\begin{itemize}
    \item \textbf{Objective 1} is to provide the research community with a structured list of open questions that must be investigated further in order for AI risk modeling to mature as a field. We aim to sharpen and add context to these questions so that they become tractable. We distinguish problems that stem from missing data, unresolved conceptual disagreement, or institutional and coordination barriers, among others, and start to identify which questions are prerequisites for progress on others.
    \item \textbf{Objective 2} is to provide practitioners and policymakers with sufficient grounding in the technical and governance issues surrounding AI risk modeling in order to engage productively with this research agenda. We have found that technical and governance questions in this domain are tightly coupled. Who builds and validates risk models, what level of transparency is required to meaningfully understand model capabilities, and what institutional mechanisms are needed to use risk models in practice are not questions that can be cleanly separated from those around model architecture and evidence integration. 
    
\end{itemize}
To advance progress toward these objectives, we hosted a workshop of twenty-two experts spanning AI safety, cybersecurity, probabilistic risk modeling, and AI governance. Insights from this session, along with subsequent analysis by the authors, were synthesized. We also provide a comparative analysis of two current methodological proposals, a scenario-based risk estimation framework \citep{touzet2025_riskmodeling,murray2025_quantitativeairisk,barrett2025_cyberrisk} and a Bayesian network risk threshold setting framework \citep{jackson2026_cyberthresholds}. We characterize their underlying assumptions, complementary strengths, and the conditions under which each is most appropriate. We then surface and organize the open questions that must be resolved for either approach, or any successor approach, to be operationalized at scale. We are concerned with AI systems rather than AI models, as most risk modeling efforts would assess the risks from the AI system as a whole (including tools, scaffolding, and interfaces) rather than the stand-alone model. Similarly, where possible, we refer to “risk models” rather than simply “models” to not confuse risk models with AI models themselves. 

There are several important things this paper does not attempt to do:
\begin{itemize}
    \item Resolve any of the open questions it identifies. We work to clarify questions and their dependencies, not to answer them.
    \item Propose a complete quantitative risk model for any specific risk domain or threat scenario.
    \item Evaluate or endorse any particular AI system, model developer, or existing governance framework.
    \item Address the full range of AI risk domains. Our focus is on catastrophic misuse risks, specifically AI-enabled threats in the cyber and CBRN (Chemical, Biological, Radiological, Nuclear) domains, and on subsets of the risk of loss of control.\footnote{Questions specific to other risk categories, such as labor displacement or algorithmic bias, are out of scope, although we note that some of the methodological questions raised here may generalize to risk modeling in those domains as well.}
    \item Provide a comprehensive survey of probabilistic risk assessment or Bayesian methods. The background on these traditions is included only to the extent necessary to situate the open questions.
\end{itemize}

The remainder of the paper is organized as follows. Section 2 reviews the five research traditions that inform current approaches to AI risk modeling: probabilistic risk assessment and safety engineering, catastrophic AI risk analysis, cybersecurity risk quantification, Bayesian networks and causal inference, and intolerable risk thresholds and red lines. Section 3 sets out the workshop context and motivation, covering current AI capability trends, the limitations of existing risk assessment methods, the case for quantitative risk modeling, and the two methodological proposals that anchor the comparative analysis. Section 4 organizes the open questions identified through the workshop and subsequent analysis along five dimensions: risk model structure, scope and granularity, evidence integration and elicitation, validation and updating, and governance and institutional design. Section 5 distills these into prioritized research areas and recommendations. Section 6 concludes.
%%%%%%%%%%%%% Section II %%%%%%%%%
\section{Background on AI Risk Modeling}
\label{sec:background}
At its core, risk modeling aims to characterize the relationship between potential hazards, the likelihood of adverse events, and the magnitude of resulting harms. In the context of AI, this requires moving beyond isolated measures of model capability to structured representations of how harms arise through sociotechnical systems. As prior work has emphasized, an effective approach to risk modeling is to combine scenario building (mapping causal pathways from hazard to harm) with risk estimation (quantifying the likelihood and severity of those pathways).

AI risk modeling sits at the intersection of several established research traditions, each of which provides partial tools for understanding and quantifying risk in complex systems. While no single tradition is sufficient to address the challenges posed by advanced AI, together they offer a foundation for developing more robust methodologies.

This section reviews five strands of literature that collectively inform current approaches to AI risk modeling: probabilistic risk assessment and safety engineering, catastrophic AI risk analysis, cybersecurity risk quantification, Bayesian networks and causal inference, and the study of intolerable risk thresholds. Each contributes distinct concepts, tools, and assumptions that shape how risks can be modeled, while also highlighting key limitations that motivate the need for integrative approaches.
%%%%%%%%%% subscetion %%%%%%%%%%%%%%
\subsection{Probabilistic Risk Assessment and Safety Engineering}
The foundational logic of quantified risk assessment originates in the nuclear and aerospace industries of the mid-twentieth century. The landmark Reactor Safety Study~\citep{nrc1975_reactorsafety}, published by the U.S. Nuclear Regulatory Commission in 1975, was the first large-scale application of fault tree analysis (FTA)\footnote{Fault tree analysis is a deductive, systematic reliability analysis method developed in the nuclear industry to evaluate the causes of system-level failures. Beginning with a predefined undesired “top event” (e.g., core damage), FTA uses Boolean logic to map how combinations introduced by lower-level component failures, human errors, and external events could lead to that outcome. The resulting fault tree structure is composed of logic gates such as AND and OR and enables both qualitative identification of critical failure pathways (minimal cut sets) and quantitative estimation of event probabilities.}  and event tree analysis (ETA)\footnote{Event tree analysis is an inductive, forward-looking reliability analysis method used to evaluate the possible outcomes following an initiating event (e.g., a loss-of-coolant accident or station blackout). Starting from the initiating event, ETA models the sequence of successes and failures of safety systems and operator actions as a branching tree structure, with each branch representing a distinct accident progression pathway. By assigning probabilities to each branch point, ETA enables both qualitative identification of key system dependencies and quantitative estimation of outcome frequencies (e.g., core damage or safe shutdown).}  to estimate the probability of catastrophic failure in a complex engineered system~\citep{nrc1975_reactorsafety}. This work demonstrated the feasibility of decomposing a complex failure into observable sub-events, assigning probabilities to each, and propagating them through a logical structure to produce a quantified risk estimate. We draw inspiration from this example in particular. Subsequent development through the aerospace and chemical process industries — including IEC 61508, the international standard for functional safety of electrical and electronic systems, and the process safety frameworks underpinning HAZOP analysis — refined these methods and embedded them into regulatory practice \citep{iec61508_1_2010}.

These methods are limited, however, in their direct application to AI. Classical probabilistic risk assessment (PRA) assumes failure modes are broadly identifiable and specifiable in advance. This works well for engineered systems with well-characterized components or operating envelopes. AI systems, particularly large language models (LLMs) that operate in open-ended environments, can fail in ways that mask the error of their outputs, and have decision logic that is opaque, making the assumption of classical PRA a poor fit without significant adaptation \citep{khlaaf2023_riskassessments}. 

To address this, Wisakanto et al. \citep{wisakanto2025_pra_ai} introduce the PRA for AI framework, which adapts the PRA tradition into a structured threat modeling methodology for AI systems. Rather than deriving point probabilities from historical failure data, the framework guides assessors through systematic hazard identification, traces causal risk pathways from system characteristics to societal harms, and produces banded risk level estimates with explicit documentation of assumptions and uncertainty. Aspect-oriented hazard analysis provides systematic coverage of the AI hazard space using a taxonomy of AI system aspects organized into four top-level categories: capabilities, domain knowledge, affordances, and impact domains. Rather than sampling the hazard space through pre-selected scenarios or commonly discussed attack chains, the framework indexes the hazard space structurally through these aspect categories, examining how each can act as a bottleneck, how aspects interact, and how competence and incompetence each generate distinct hazards. Risk pathway modeling traces causal progressions from source aspects to terminal aspects using bidirectional analysis, complementing forward propagation with backward tracing from impact domains to enabling system characteristics. Uncertainty is managed through standardized intensity rubrics, classification heuristics, and explicit uncertainty tracing protocols. These results are then synthesized into a risk report card aggregating estimates across aspect categories. This approach establishes a robust and defensible PRA method for AI on which both scenario-based and probabilistic risk modeling approaches can operate.

%%%%%%%%%% subsection %%%%%%%%
\subsection{AI Catastrophic Risk Analysis}

A second tradition originates in the philosophical and technical literature on catastrophic risks posed by advanced AI systems. Early foundational work established the conceptual case that sufficiently capable AI systems could pose qualitatively different risks from those associated with prior technologies, including the potential for large-scale or irreversible harm \citep{russell2019_humancompatible,ord2020_precipice}. This body of work has emphasized the importance of long-term and low-probability, high-impact risks, often under conditions of deep uncertainty and limited empirical precedent.

More recent contributions have begun to move this tradition toward greater empirical and operational grounding. Efforts such as those by \citet{hendrycks2023_catastrophicrisks} and \citet{slattery2024_airisk_arxiv} articulate concrete threat models for advanced AI systems, including misuse risks (e.g., AI-enabled cyber or biological attacks) and risks arising from loss of control or misalignment. These works aim to translate abstract concerns into more structured descriptions of how harms could materialize, often by identifying key actors, capabilities, and pathways.

Most existing work in this tradition remains primarily qualitative. While scenario development has become increasingly sophisticated, there are limited methods for assigning probabilities to these scenarios or for comparing their relative importance in a systematic way. As a result, it is difficult to use these analyses directly for decision-making tasks such as prioritizing mitigations, setting risk thresholds, or evaluating trade-offs between different policy options.

Another key challenge lies in handling uncertainty and disagreement. Catastrophic AI risks often involve deep epistemic uncertainty, where both the structure of scenarios and the values of key parameters are contested. Existing approaches typically represent this uncertainty implicitly, through narrative discussion or the exploration of multiple scenarios, rather than through explicit probabilistic modeling.

%%%%%%%%%%%%%%% subsection %%%%%%%%%
\subsection{Cybersecurity Risk Quantification}

The existing body of work on quantitative risk assessment within the cybersecurity field has also been a source of inspiration. The Factor Analysis of Information Risk (FAIR) framework \citep{fairinstitute_whatisfair} is the most mature prior attempt to bring actuarial-style quantification to cyber risk. FAIR decomposes cyber risk into a hierarchy of factors (threat event frequency, vulnerability, and loss magnitude) and provides a structured method for estimating probability distributions over each, enabling Monte Carlo aggregation to an overall risk estimate. The Common Vulnerability Scoring System (CVSS) \citep{first_cvss} provides a complementary resource, with a standardized vocabulary for scoring the severity of individual software vulnerabilities.\footnote{There is now also an AI-specific resource, the AI Vulnerabilities Scoring System (AIVSS) \citep{owasp_aivss}.} Threat modeling methodologies, including STRIDE \citep{microsoft2009_stride} and PASTA \citep{ucedavelez2015_riskcentric}, provide structured vocabularies for decomposing attacker behavior into observable, assessable steps. For understanding the landscape of cyber-specific threats, frameworks like MITRE ATT\&CK \citep{mitre_attack} provide insight into real-world kill chains. However, these frameworks were designed for human-driven threats with relatively stable characteristics. They do not account for AI systems as autonomous threat actors, cannot model rapid capability evolution that shifts the exploitability of vulnerabilities, and lack the causal structure needed to trace how specific AI capabilities enable particular attack pathways.

%%%%%%%%%%%% Subsection %%%%%%%%%%%%
\subsection{Bayesian Networks and Causal Inference}
Bayesian networks (BNs) were formalized by Judea Pearl as a language for probabilistic reasoning \citep{pearl2014probabilistic}. Through the representation of conditional dependencies and the subsequent elaboration of do-calculus (a mathematical framework for predicting the effects of deliberate interventions), Pearl provided the tools necessary to model systems where variables have structured causal relationships rather than varying independently \citep{pearl2009_causality}. This methodology has since been refined across various high-stakes domains, from medical diagnosis and fault detection to actuarial science and intelligence analysis \citep{lauritzen1988local,fenton2018_bayesiannetworks}.

For AI risk modeling, BNs offer two primary advantages. First, they make causal structure explicit, allowing modelers to represent the conditional logic of, for example, a multi-stage cyberattack where the success of later stages is strictly contingent on earlier ones. In contrast, simpler models, such as some FAIR-style models, often treat variables as independent or weakly correlated, missing the conditional dependencies that BNs can represent. Second, they provide a framework for synthesizing evidence. Bayesian updating allows risk estimates to be refined as new data arrives from capability benchmarks, red-team exercises, or observed incidents. Despite these strengths, BNs face scaling limitations, both in the computational complexity of performing inference and the parameterization of the network. The difficulty of eliciting well-calibrated parameter estimates from domain experts is a further challenge \citep{koller2009_pgm,druzdzel2013elicitation}.

%%%%%%%%%%%%%% subsection %%%%%%%
\subsection{Intolerable Risk Thresholds and Red Lines}

This emerging area in AI safety discourse draws from international humanitarian law, arms control, and the ethics of absolute constraints. The prohibition frameworks governing chemical weapons, biological weapons, and anti-personnel landmines all make the claim that some risks are categorically intolerable, requiring prohibition rather than managed mitigation \citep{unoda_bwc,cwc1993,ottawa1997_minebantreaty}. Within the life sciences, the Dual-Use Research of Concern (DURC) framework, articulated in the Fink Report, represents the most technically mature attempt to institutionalize threshold-based governance over capabilities that could enable mass harm \citep{national2004biotechnology}.

In the AI domain, these efforts began with the Seoul Ministerial Statement \citep{seoul_ministerial_2024} and Frontier AI Safety Commitments \citep{frontier_ai_commitments_2024} that were signed by 27 nations and the EU, and 16 AI developers, respectively at the Seoul AI Safety Summit \citep{seoul_declaration_2024}. Companies have since published Frontier Safety Frameworks, representing industry-led attempts to define specific capability thresholds that would trigger mandatory mitigation actions, such as deployment pauses or appropriate information security levels \citep{fmf2025_risktaxonomy,metr2025_commonelements,stelling2025evaluating}. Accompanying literature has also identified principled approaches to identifying thresholds, red lines and other best practices across organizations to strengthen proactive risk-management efforts \citep{raman2025intolerable,tfs2025_redlines,ziosi2025_safetyframeworks}. Other, broader, civil society and internationally endorsed work, such as the campaign for the Global Call for AI Red Lines \citep{redlines2025_globalcall}, has also gained traction with signatories from across the world appealing for clear and verifiable thresholds for universally unacceptable risks.

However, these threshold definitions are largely qualitative and/or capability-based. Phrases such as “meaningful uplift,” when referring to biological weapon creation, correctly identifies a risk area of concern, but lacks the quantitative operationalization needed for consistent enforcement. Emerging regulatory action underscores this need for actionable criteria, and this is a central motivation for emerging methodologies aimed at moving from qualitative descriptors to probabilistic measures \citep{jackson2026_cyberthresholds}.\footnote{The EU AI Act introduces the legal category of “unacceptable risk,” prohibiting certain AI systems expected to produce systemic harms or violate fundamental rights, thereby codifying the principle that some AI applications must be categorically disallowed rather than merely mitigated (Regulation 2024/1689) \citep{eu_ai_act_2024}. Recent U.S. regulatory initiatives similarly move toward formalizing threshold-based governance. For example, California Senate Bill 53 requires frontier-model developers to define and assess “catastrophic risk” thresholds — defined as risks that could plausibly contribute to events causing more than 50 deaths or over \$1 billion in damages — and to publish risk assessments and mitigation protocols tied to those thresholds \citep{ca_sb53_2025}. Likewise, New York’s RAISE Act obligates large AI developers to establish and disclose safety frameworks addressing severe harms such as automated crime or AI-assisted bioweapon development, along with mandatory reporting of critical safety incidents \citep{ny_raise_act_2025}. Together, these efforts reflect a broader shift in AI governance toward defining enforceable “red lines” grounded in measurable risk thresholds rather than purely qualitative capability descriptions.} 

%%%%%%%%%%% Section Workshop %%%%%%%%%%%%

\section{Workshop Context and Motivation}
\label{sec:Workshop}

\subsection{Growing AI Capabilities and Associated Risks}
General scaling of LLMs has led to the rapid advancement of capabilities across an array of economically valuable tasks. These capabilities include information synthesis, sophisticated code generation, vulnerability discovery, social engineering, and the autonomous execution of complex multi-step planning and reasoning. While such capabilities promise significant societal utility, they also simultaneously introduce or exacerbate catastrophic risks, including cybersecurity, biological security, and large-scale manipulation.

A defining characteristic of these risks is that they are not inherent to the capabilities themselves, but emerge through the dynamic interplay between model capabilities, threat actor behavior, and the vulnerabilities of real-world systems. Model capabilities should be viewed as risk sources, not risks in themselves. This distinction is important because improvements in performance on capability benchmarks or task-specific metrics do not translate directly into realized societal harm; instead, they operate through intricate, multi-stage causal pathways involving human agents, specific targets, and broader environmental contingencies.

Emerging evidence within the cybersecurity domain illustrates this dynamic. Quantitative modeling of AI-enabled offensive cyber operations indicates that AI systems may significantly enhance the efficacy and scale of attacks and broaden the spectrum of viable targets. These effects are driven by several distinct mechanisms: improving the success probability of individual attack chain steps, increasing attack throughput per unit of time, lowering the technical barriers to entry for novice actors, and facilitating complex coordination across disparate attack stages. Critically, these effects remain heterogeneous across threat scenarios, with varied combinations of factors driving the aggregate risk profile depending on the specific operational context.

More broadly, the AI risk landscape is defined by three characteristics that distinguish it from the failure modes encountered in traditional safety-critical industries:

\begin{itemize}
    \item Open-ended capability space: The generalizability of AI systems enables the emergence of novel and difficult-to-enumerate failure modes that defy traditional closed-world assumptions.
    \item Rapid capability scaling: Continuous improvements in underlying architectures can suddenly shift the probability distributions of associated risks.
    \item Sociotechnical dependence: Risk outcomes are highly contingent upon factors such as human behavior and the specific conditions of deployment.
\end{itemize}
Collectively, these features suggest that the interplay between AI capabilities and societal risk is not straightforward but rather highly context-dependent. Consequently, robust risk assessment necessitates a transition from capability-centric evaluations toward modeling frameworks that connect capabilities, deployment context, and other relevant inputs to the harms that are realized.

%%%%%%%%%%%%% subsection %%%%%%%%%%%%%
\subsection{Limitations of Current Risk Assessment Methods}

Despite an increased focus on AI-related hazards, current AI risk assessment practices remain methodologically insufficient to capture the underlying causal structure or the full magnitude of these risks. Existing approaches generally coalesce into three categories: capability-centric evaluations, qualitative safety cases, and fragmented quantitative studies. While each provides partial visibility, they suffer from significant limitations when applied in isolation to frontier AI systems.

First, capability-centric evaluations do not constitute direct measures of risk. Frontier AI safety frameworks often rely on static benchmarks to trigger mitigation responses once predefined thresholds are reached. However, these methods assess only the existence of capabilities rather than the probability or severity of actual resulting harms to society. Consequently, they fail to provide answers regarding the absolute risk of a system or whether proposed mitigations are commensurate with the threat.

In addition to this uncertain external validity, benchmarks are imperfect proxies in other ways. Their performance may conflate disparate underlying capabilities and often have issues with construct validity and reproducibility \citep{eriksson2025can}. Lately, there have also been increasing issues with benchmark saturation \citep{akhtar2026ai}. This introduces significant epistemic ambiguity into the assessment process, limiting its utility for high-stakes decision-making.

Further, current methodologies that examine isolated capabilities fail to capture the many-layered interactions between factors. Risks in the cyber and CBRN domains are realized through a multi-faceted sequence, involving threat actor behavior and incentives as well as targets with very differing levels of defenses. Loss of control or autonomy risk scenarios may involve systems purposefully evading safeguards or pursuing misaligned objectives. Agentic risk scenarios often include an interplay between multiple agents and cascading risk failures. Omitting analysis of this interplay creates critical blind spots and can lead to both under- and overestimating the level of risk.

Safety cases offer structured arguments for safety but are not intended to provide exhaustive coverage of the risk space or to generate quantified estimates of likelihood and impact. Furthermore, they are also plagued with the same limitations in their verifiability \citep{khlaaf2025safety}. Rather than serving as a substitute for formalized, probabilistic risk modeling, safety cases should be understood as a complementary framework that incorporates risk modeling outputs as evidence. Quantitative risk models can provide the causal pathways, probability estimates, and severity assessments that populate the evidence layer of a safety case, while the safety case structure organizes these inputs into a coherent argument for deployment decisions.

Third, qualitative scenario building and quantitative estimation are often decoupled. In current practice, scenario analysis and statistical estimation are frequently conducted as independent workstreams. This leads to quantification driven by data availability rather than causal relevance, and scenario analysis that lacks decision-relevant probabilistic outputs. As highlighted in earlier work, estimation without a structured scenario framework cannot produce a coherent risk profile, while scenarios without quantification are insufficient for setting defensible thresholds or navigating governance trade-offs.

Fourth, traditional approaches struggle to account for dynamic feedback loops and interacting effects. Advanced AI systems can fundamentally shift risk distributions by enabling novel attack strategies, altering the offense-defense balance, and creating interactions between capabilities and countermeasures. Methodologies developed for relatively static, closed-world systems are poorly suited to capturing these shifting sociotechnical dynamics.

Collectively, these limitations imply that existing practices are insufficient to support essential risk management functions, including:
\begin{itemize}
    \item Comparative analysis of risks across disparate systems or domains.
    \item Quantitative evaluation of mitigation effectiveness.
    \item Establishing and enforcing rigorous, measurable risk thresholds.
    \item Prioritizing evaluation resources and monitoring efforts.
\end{itemize}
Addressing these gaps motivates a methodological transition toward frameworks that explicitly represent the causal mechanisms through which risks emerge and evolve.

%%%%%%%%%%% subsection %%%%%%%%%%%%
\subsection{The Need for Quantitative Risk Modeling Methods}

To address these systemic methodological gaps, there is a critical need for quantitative risk modeling approaches that map AI capabilities to societal harms through structured causal representations. These frameworks seek to bridge the epistemological distance between hazards (latent model capabilities) and harm (realized economic or physical damage) by explicitly modeling the intermediate causal pathways that link them.

At a high level, this quantitative transition integrates two primary components:

\begin{enumerate}
    \item {Scenario building:} The definition of concrete, plausible causal pathways from initial hazard to realized harm.
    \item {Risk estimation:} The assignment of probability and impact distributions to each discrete step within these pathways.
\end{enumerate}

This approach enables the construction of models capable of producing decision-relevant outputs, such as the probability of a system exceeding a specific harm threshold within a defined time horizon.

This structural decomposition provides a framework for synthesizing heterogeneous evidence (e.g., capability benchmarks, structured expert elicitation, historical incident reports, and red-teaming data) into a unified probabilistic model.

Furthermore, quantitative risk modeling facilitates several governance functions:

\begin{itemize}
    \item Enables cross-domain comparative analysis by situating AI-enabled risks relative to established threats in other domains.
    \item Supports threshold-based governance through measurable red lines for intolerable risk.
    \item Provides uncertainty quantification by identifying the primary drivers of epistemic uncertainty and expert disagreement.
    \item Allows for updating as new evidence emerges, incorporating new threat intelligence and capability data as they surface.
\end{itemize}

Significantly, the process of model construction provides value prior to full quantification by clarifying underlying assumptions, surfacing evidence gaps, and identifying critical bottlenecks in threat pathways.

%%%%%%%%%%%% subsection %%%%%%%%%%%%
\subsection{Current Methodological Proposals}
A small number of research publications have emerged as a methodological foundation for modeling AI-enabled risks. This section outlines two methodological approaches — a scenario-based risk estimation framework and a Bayesian network for risk threshold setting framework — and related research that supports them.

%%%%%%%%% sub-subsection %%%%%%%%%%%
\subsubsection{Scenario-Based Risk Estimation Framework}

Murray et al. (2025) \citep{murray2025_quantitativeairisk} introduce a quantitative risk modeling approach through a structured six-step process that integrates scenario building with parameterized risk estimation. The framework decomposes risks along key dimensions (such as actor-target-vector taxonomies) to select representative scenarios, then models each scenario as a combination of components: the frequency of initiating events, the probability of successful completion of the event chain, and the magnitude of the resulting harm. Baseline risk is established using historical data, then marginal (“uplift”) risk from LLM capabilities is estimated by mapping Key Risk Indicators (such as benchmarks) to individual risk parameters through structured expert elicitation. For scalability, the methodology includes experimenting with both human experts and LLM-simulated “experts”. The methodology aggregates leaf-level probabilities by fitting Beta distributions to expert estimates and propagating uncertainty through Monte Carlo simulation. This produces actionable quantitative outputs (e.g., uncertainty intervals on annual expected damages) that can enable informed decision-making, used to prioritize mitigation efforts by identifying bottleneck steps in attack pathways or inform thresholds. 

An associated paper, by Barrett et al.\citep{barrett2025_cyberrisk}, applies the methodology to AI-enabled cyber offense, developing a suite of risk models using the MITRE ATT\&CK framework as a taxonomy for step-level decomposition. These models assess AI uplift across four dimensions: attacker population size, attack frequency per actor, step-level success probabilities, and harm magnitude. Cybench and BountyBench (two common cyber risk benchmarks) serve as the primary Key Risk Indicators, with benchmark-to-parameter mappings estimated through both human and LLM-simulated expert elicitation. The companion analysis finds systematic uplift effects across attack types, though the mechanisms driving uplift vary across scenarios — some are driven by capability expansion, others by throughput gains or lowered barriers to entry.

In an additional accompanying paper, Touzet et al.~\citep{touzet2025_riskmodeling} provide the theoretical and comparative foundation for AI risk modeling, examining classical risk modeling techniques — including fault tree analysis, event tree analysis, FMEA, STPA, and Bayesian networks — and assessing how each can be adapted for advanced AI systems. This work argues for an approach that combines deterministic guarantees for categorically unacceptable events with probabilistic assessment across the broader risk landscape. It identifies fragmentation in current practice across capability benchmarks, safety cases, and partial quantitative studies as the central obstacle to progress.

%%%%%%% sub-subsection %%%%%%%%%%%%
\subsubsection{Bayesian Network for Risk Threshold Setting Framework}

Jackson et al.~\citep{jackson2026_cyberthresholds} propose an approach that prioritizes system-level causal structure over scenario-level operational specificity. Rather than decomposing individual attack chains, this framework uses Bayesian networks to represent the complex, interdependent variables that jointly determine whether an AI-enabled cyber threat crosses a governance-relevant risk threshold. A central design objective is to operationalize such thresholds as defensible, measurable probabilistic quantities that can trigger specific governance responses, such as deployment pauses or mandatory audits. The Bayesian network structure supports explicit modeling of the offense-defense balance by incorporating mitigating controls and defensive capabilities as nodes within the causal graph. Different forms of uncertainty can be separately represented and weighted, enabling more transparent and auditable risk claims.

The framework also acknowledges several limitations. Since standard Bayesian networks are directed acyclic graphs (DAGs), they cannot directly represent recursive feedback loops without additional extensions, which limits their suitability for modeling dynamically interacting systems. Additionally, as network complexity increases, the conditional probability tables associated with each node can grow exponentially. Finally, eliciting well-calibrated parameter estimates in a domain with sparse historical data introduces significant epistemic uncertainty and requires ongoing validation against emerging threat intelligence. 

%%%%%%%%%%%%%% section open Qs %%%%%%%
\section{Open Questions}
\label{sec:OpenQuestions}

Recent approaches to AI risk modeling have laid some of the groundwork, but the field remains nascent. Significant gaps persist across technical, methodological, and institutional dimensions, limiting the reliability, scalability, and practical usefulness of existing models. Notably, many of the core design choices involved in constructing and applying AI risk models remain unresolved, creating a bottleneck to further progress.

This section synthesizes the key open questions identified through several literature reviews and the discussion at our expert workshop. These questions span multiple layers of the modeling process, including: model architecture, scope and granularity, evidence integration and elicitation, validation and updating, and the governance structures required to support effective use. In many cases, these challenges are interdependent, decisions about model structure influence what data is needed, limitations in available evidence constrain model fidelity, and governance choices shape incentives for model development, validation, and disclosure.

Rather than attempting to resolve these issues, we aim to clarify and organize them in a way that makes them more tractable for future research. In particular, we distinguish between questions driven by data limitations, conceptual uncertainty, and institutional constraints, and highlight where progress on one set of questions may be a prerequisite for others. Together, these open questions define a research agenda for advancing AI risk modeling toward a more robust and operational discipline.

%%%%%%%%%%% subsection %%%%%%%%%
\subsection{Risk Model Structure}

\subsubsection{Data Types and Challenges}

Since risk modeling serves as a bridge between AI capability assessments and real-world risk, it is essential for risk models to use high-quality data as inputs in order to make the outputs meaningful. Broadly speaking, this data can fall into the following categories:
\begin{itemize}
    \item \textbf{Experimental:} Experimental/simulated data/evaluations and benchmarks
    \item \textbf{Historical:} Empirical/observational data/incident reports 
    \item \textbf{Expert:} Expert-elicited parameter estimates/structured expert judgments 
\end{itemize}
All three types have both advantages and drawbacks and, in the future, we believe it will be desirable to combine them. However, principled ways of selecting and aggregating these sources of information remain an open problem.

Benchmark and evaluation data are widely recognized as insufficient for risk modeling purposes. They are most useful as indicators of whether a capability exists, not as measures of how likely or consequential its use in practice would be \citep{eriksson2025can,mcintosh2025inadequacies,weidinger2025toward}. Workshop participants raised concerns about benchmark saturation, evaluation awareness, the risk of gaming, under-reporting, and the lack of standardized reporting templates that would make benchmark outputs comparable across models and organizations. There was broad agreement that benchmarks need to be tailored to specific risk scenarios rather than used generically, and that developing better standards for evaluation reporting was seen as a meaningful near-term contribution. It was noted that there is a positive move toward the use of open-ended evaluations that are closer to real-world scenarios, such as the UK AISI’s cyber ranges  \citep{aisi2026_mythoscyber}. 

Historical data — for example, incident tracking portals (e.g., the MIT AI Incident Tracker) \citep{mit_airisk_incidenttracker}, threat intelligence reports, or unstructured sources on the internet — can be useful as inputs to risk models. Databases collecting historical incidents and data already exist and may be incorporated into risk modeling methodologies. However, workshop participants raised concerns around reporting biases, such as selection effects and developers highlighting positive examples and not failures. Additionally, historical data is often available at a less granular level, providing a signal that an incident occurred, but without detail on the specific tools or capabilities displayed by AI systems. Determining the extent to which a given incident can be attributed to AI systems as opposed to humans is also challenging. For example, in the context of cybersecurity, common vulnerability tracking portals typically do not record AI’s contributions, even if it is confirmed by other sources.

Historical data is useful for informing models initially and can also be used as a tool for the backtesting of risk models to calibrate their predictions against past values. Workshop participants noted that there can be large variations in reporting across sources, in particular around damage and harm estimates, making backtesting challenging even if relevant historical data is available. Additionally, obtaining ground-truth data about activities such as cyber or CBRN attacks executed without assistance from frontier LLMs will become increasingly difficult as AI gets integrated into operations in non-obvious ways. As a result, estimates of AI-driven uplift will not have a reliable baseline to start from and may only be possible by actually performing uplift studies with recruited participants.

Expert judgment is a viable option for obtaining data relevant to risk modeling and has been used across a variety of contexts ranging from nuclear safety to terrorism studies. Nonetheless, collecting it remains time and resource-intensive, as well as susceptible to mental biases. As such, it typically needs to be carried out in structured environments under the supervision of an experienced facilitator. It is also necessarily reliant on the aforementioned source of empirical data, such as evaluations or real-life incidents, since without any prior data to calibrate on, experts struggle to give informed judgments.

%%%%%%%%%%% sub-subsection %%%%%%%%%%%
\subsubsection{Model Architecture}
Risk model architecture refers to the structural choices made before any specific parameter is estimated. Important decisions need to be made on how the causal pathways from hazard to harm are represented, what variables and dependencies are included, and how aggregate risk is computed from the individual components. These choices are difficult to revisit once decided on and so they merit careful consideration early in the process. Workshop participants identified several open questions in this space. \\

\textit{Appropriate structural representation} \\
The first issue concerns the appropriate structural representation. Multiple options exist, for example: fault trees, event trees, chain event graphs, Bayesian networks, and influence diagrams. Each carries different assumptions about how harms unfold and how evidence is combined, and there is currently no consensus on which formalism is best suited to which class of AI risk. If multiple groups model the same risk scenario using different architectures, comparing the outputs becomes a challenge on its own.\\

\textit{Parameter parsimony} \\
The second problem concerns parameter parsimony. As the number of parameters increases, the number of pairwise correlations between them grows quadratically. A model with five parameters has ten such pairs; a model with ten parameters has forty-five. In practice, most current models simply assume parameter independence, but this assumption is rarely justified. For instance, LLM capabilities are likely highly correlated across the various steps of risk scenarios. Capabilities tend to move in tandem, e.g., a model that has valuable capabilities for malware creation is also likely to have valuable capabilities for evasion detection. Explicit modeling of parameter dependencies remains underdeveloped and the cost of doing it well provides a strong practical argument for parsimony. In some cases, modeling interdependencies explicitly might introduce new hyperparameters whose uncertainty defeats the purpose of introducing them in the first place. \\

\textit{Static nature of risk models} \\
The third issue concerns the static nature of risk models. Most current risk models represent a snapshot of activities — a likely attack pathway against an assumed defender posture. The reality, however, is much more dynamic. Defenders adapt, mitigations are deployed, and attackers shift tactics in response to a changing threat and capability landscape. This is true both within particular events (i.e., attacks might last hours, days, or weeks and involve a constant cat-and-mouse game that is difficult for modelers to capture) and over time (both offensive and defensive practices change from month to month). Given the incredibly fast pace of AI progress and the adaptability of AI systems to novel scenarios, simply modeling one representative attack chain might not be enough to cover the annual risk landscape, let alone a multi-year forecast. Embedding these dynamics directly into the architecture (e.g., through time-indexed nodes, defender response functions, or explicit feedback between attack and defense) is technically feasible, but substantially increases the parameter burden and uncertainty levels. The alternative is to keep the architecture static, but update or produce new risk models frequently, or to consider other modeling approaches, such as agent-based modeling — a computational approach that simulates the actions and interactions of autonomous agents within a defined environment. However, workshop participants expressed concerns that the ecosystems of risk modeling, as well as AI evaluations, are both under-resourced and will struggle to keep up with evolving or novel risks.

A final question concerns the extent to which models can be built from shared structures or reusable components. As multiple groups develop models across different domains and scenarios, it may be possible to standardize elements such as common subgraphs, taxonomies of attacker steps, or recurring risk factors. Shared representations of attacker behavior, system vulnerabilities, or propagation dynamics could serve as building blocks across models, improving comparability and reducing duplication of effort. However, this requires agreement on common abstractions and on mechanisms for maintaining and updating shared components over time. The MITRE ATT\&CK \citep{mitre_attack} framework is a good starting point in the cybersecurity context, but at the moment, no equivalent framework exists for domains such as CBRN, harmful manipulation, or loss of control.

%%%%%%%%% sub-subsection %%%%%%%%%%%%

\subsubsection{Using Multiple Models}
A point of emerging consensus in the workshop was that the field should expect to need multiple models rather than a single unified risk model, and that this proliferation is a feature rather than a problem to be solved. There are two distinct motivations for moving in this direction.

First, answering different governance questions may require different model architectures. The question of whether a model has crossed a risk threshold calls for a model optimized for clarity around the threshold boundary, whereas the question of which mitigations are most effective calls for a model that accurately captures the causal pathways through which interventions reduce risk. Trying to answer distinct governance questions with one model increases the risk of introducing systematic distortions.

Second, ensemble approaches — maintaining multiple models with slightly different structural assumptions or parameters and comparing their outputs — may be the most principled and defensible path forward. The deep uncertainty that characterizes this domain requires a pragmatic approach that balances the immense value of using risk models with a realistic analysis of their limitations. A statistical approach that averages the results of multiple models would allow us to continue treating uncertainty as a key characteristic rather than a flaw of these methods. This is analogous to the approach taken in climate science and infectious disease epidemiology, where Bayesian model averaging (BMA) is used to combine predictive distributions from multiple models, weighting each by its posterior probability of predictive skill~\citep{hoeting1999_bma,raftery2005_weatherbma}. In weather forecasting, BMA has been shown to produce calibrated probabilistic outputs and sharper prediction intervals than raw ensembles~\citep{raftery2005_weatherbma}. Such ensemble approaches can form a basis for communicating both estimates and uncertainty ranges to policymakers.

When models built on different assumptions produce convergent risk estimates, confidence in those estimates increases; when they diverge, the divergence itself is informative, identifying the structural assumptions that are most consequential for the final risk picture. Rather than seeking a single “best” model, this perspective treats different models (that could vary in structure, assumptions, or level of granularity) as complementary sources of insight. Combining multiple imperfect models can help triangulate estimates, identify robust directional signals, and expose sources of disagreement. Critically, alignment across models may matter more for decision-making than absolute accuracy. Two models could produce significantly different raw estimates for annual expected damage yet still agree on the relative value of interventions (e.g., intervention A > B > C). Such ordinal agreement is a positive signal even when cardinal estimates diverge. In this view, the use of multiple weak or approximate methods is not merely a fallback in the absence of better data, but a core methodological strategy for improving robustness in highly uncertain domains. This, however, does raise further questions about how to systematically combine outputs across models and how to communicate uncertainty when different approaches yield divergent results.

An important implication of this perspective is that different applications of risk models may require different types and strengths of evidence. For example, using models to set or enforce risk thresholds may require conservative assumptions and upper-bound estimates, whereas applications such as prioritizing mitigations or identifying key uncertainties may be adequately supported by lower-bound or directional estimates. This suggests that the evaluation of model quality cannot be divorced from its intended use: the same model may be considered sufficient for one purpose and inadequate for another. Clarifying these distinctions is essential for aligning modeling approaches with decision-making needs.

%%%%%%%%% subsection %%%%%%%%%%%

\subsection{Model Scope and Granularity}
An open question in AI risk modeling, and risk modeling more broadly, is how to balance scope and granularity. This requires determining how comprehensively a model should cover the “risk universe” versus how finely it should delineate causal architecture within discrete risk pathways.

While there is a general consensus that for many types of questions, models should represent causal sequences from hazard to harm through some type of structured decomposition (e.g., event chains or interacting risk factors), significant disagreement remains regarding the necessary degrees of resolution. This introduces a fundamental tension between prioritizing comprehensiveness and representational fidelity through complexity, or favoring tractability and structural robustness through more parsimonious, aggregated approaches. \\

\subsubsection{Coverage of Risk Universe} 
One critical dimension involves the extent to which models can realistically map the full spectrum of potential risk scenarios. In practice, the risk landscape for frontier AI is vast, forcing any modeling effort to prioritize a narrow subset of scenarios. This raises significant concerns regarding completeness; if the selected scenarios capture only a fraction of the total risk space, the resulting outputs may systematically underestimate the aggregate threat.

As underscored during the expert workshop, this suggests that many contemporary methodologies may essentially function as lower bound estimates of risk. While such indices remain valuable for identifying dominant pathways or comparing interventions, they are less suited for governance functions requiring threshold-setting or safety assurance, where upper-bound or worst-case characterizations are key.

This challenge is exacerbated by the lack of objective criteria for what constitutes a “sufficient” scenario set. Even within cybersecurity, a relatively mature domain with established taxonomies, coverage remains partial. In less empirically grounded areas, such as loss-of-control scenarios, the absence of historical analogues makes comprehensive mapping even more elusive. \\

\subsubsection{Granularity of Individual Models} 
A second dimension concerns the depth of detail required to model individual scenarios. Highly granular architectures (such as intricate event trees or expansive Bayesian networks) can theoretically capture subtle dependencies and provide precise estimates. However, these models require significantly more robust data and stronger structural assumptions for parameterization.

Workshop participants expressed skepticism regarding whether increased complexity is justified given current data constraints. In many risk domains, available evidence is sparse, noisy, or prone to bias, and expert elicitation may struggle to accurately reflect real-world dynamics. Under such conditions, heightened granularity may provide a false sense of accuracy, yielding results that are hypersensitive to poorly grounded assumptions.

This implies a distinct trade-off: increased model complexity may enhance theoretical fidelity, but often decreases empirical robustness, particularly under deep uncertainty. Simplified models, which rely on coarser heuristics or higher-level abstractions, may offer more stable and interpretable insights for decision-making.

A recurring theme across both dimensions is the imperative to explicitly quantify uncertainty. When data quality is poor, increasing complexity can amplify epistemic noise. Consequently, participants highlighted the utility of aggregating multiple weak signals, including diverse data sources and heterogeneous elicitation methods, to generate at least order-of-magnitude estimates.

%%%%%%%%% subsection %%%%%%%%%%%%%%%%
\subsection{Evidence Integration and Elicitation}

Model reliability is strongly tied to evidence reliability. This section addresses the open questions surrounding how evidence is gathered, weighted, and ultimately translated into model parameters.

A central challenge in integrating empirical evidence into AI risk models is an operationalization gap between what existing evaluations measure and what risk models require. Most current evaluations are designed for reproducibility, benchmarking, and comparability across models, often under controlled and simplified conditions. In contrast, risk models require inputs that reflect real-world dynamics, including adaptive adversaries, heterogeneous contexts, and multi-step interactions between system capabilities and human actors. As a result, evaluation outputs do not map cleanly onto model parameters, and direct translation from benchmark performance to real-world risk remains difficult.

More broadly, the increasing reliance on expert elicitation in AI risk modeling can be understood as a response to limitations in current evaluation methods. Existing benchmarks and evaluations often fail to capture the full complexity of real-world risk, particularly in multi-step or adversarial settings. As a result, elicitation is frequently used to bridge the gap between what can be measured and what must be estimated. This suggests that elicitation is, in part, a compensatory mechanism for shortcomings in evaluation design. While elicitation provides flexibility and can incorporate contextual knowledge, over-reliance on it risks shifting the burden of uncertainty onto subjective judgment. This dynamic underscores the importance of improving evaluation methods alongside refining elicitation techniques.

%%%%%%% sub-subsection %%%%%%%%%%%%%%%%%%
\subsubsection{Elicitation Methods}
Expert elicitation refers to formalized methods for extracting and aggregating expert judgments to produce probabilistic estimates and to quantify uncertainties. This process is often both costly and time-consuming. However, it remains an invaluable resource for the construction of risk models, especially in domains where historical and experimental data are limited. The workshop’s participants identified several options for who or what can perform this translation: (1) human domain experts, (2) human forecasting experts, and (3) LLM-simulated experts. Each approach presents distinct tradeoffs across scalability and bias. 

Human domain experts offer the highest validity levels, but are scarce. Workshop participants noted that experts who have both deep domain knowledge and sufficient context on AI development are even more rare, but that experts who have both are necessary to obtain meaningful estimates. We note that a strong operational understanding of how AI systems function in practice and their potential impact on specific domains is information that is likely to become more widespread over time.

Human forecasting experts were raised as a potential partial substitute. Workshop participants noted that some well-calibrated generalist forecasters or superforecasters could perform competitively with domain specialists in certain estimation tasks \citep{mellers2015psychology}. However, concerns were raised about whether this would apply to the highly technical and novel domain knowledge needed for AI risk assessment. The question of whether superforecasters can usefully estimate, for example, the probability that a given LLM capability translates into meaningful uplift for a specific step in a cyberattack remains open.

LLM-simulated experts are a third option, as a potential replacement or supplement to human experts. While some workshop participants expressed concerns about this approach, given its novelty and lack of empirical validity, there was also a strong desire to assess its appropriateness as a substitute for human-based expert elicitation. The reasons were a direct continuation of the discussion on the limitations of human experts, i.e., a lack of expertise spanning risk domains and AI development, the scarcity of experts overall, and the cost of these elicitation procedures. LLM-simulated experts have been explored in both \citet{murray2025_quantitativeairisk} and \citet{barrett2025_cyberrisk}. Recent work demonstrating that LLM-simulated experts can produce risk estimates highly correlated with those of human experts or ground-truth baselines was noted as promising. 

Notably, Capstick et al. \citep{capstick2024autoelicit} showed that, using their AutoElicit framework, LLMs can automatically generate prior distributions for Bayesian linear predictive models. Experts can further improve estimates by providing domain knowledge in natural language, which the LLM can then incorporate. This has the potential to make Bayesian modeling more accessible in low-resource or data-scarce settings, a central problem raised by  \citet{jackson2026_cyberthresholds} in modeling novel AI-enabled cyber threats. However, important methodological gaps still exist. AutoElicit was designed and validated for structured predictive tasks such as clinical classification, and the nodes related to AI risk might be considerably more ambiguous and context-dependent.  \citet{selby2025had} also explored LLM elicitation of prior distributions for Bayesian models, finding that while LLMs can produce informative priors, they are unreliable in certain respects. They found bias toward well-represented contexts, inconsistency across models, and difficulty validating without ground truth. 

\citet{quarks2025_llmexpertjudgement} demonstrated that LLM elicitation can produce internally coherent and directionally plausible estimates on cybersecurity domain-relevant risks. There is still a need, however, to demonstrate that these estimates are accurate in an absolute sense. Calibration against ground truth remains a key unresolved challenge, and the divergence between human expert groups on these topics means that even the human-elicited baseline is uncertain. Recent benchmarking efforts like ForecastBench \citep{karger2025forecastbench} provide infrastructure for comparing forecasting performance across humans and AI systems, though they have not yet targeted the domain-specific technical questions most relevant to AI risk. However, efforts such as these could be tailored to provide better elicitation on highly specific risks. 

Overall, while LLMs show promise as replacements or supplemental experts, their outputs are heavily shaped by the composition of their training data. As such, LLM-elicited estimates are simply a reflection of the data and refinement provided during training, and are not free from bias or potential manipulation. LLM-simulated experts should be used in a structured process, similar to that used for human experts (e.g., Delphi processes). It was also demonstrated that prompting, such as giving LLMs expert personas, had a positive effect on outputs \citep{selby2025had}. What makes a good probability distribution is ultimately subjective, but methods for quantifying the utility of synthetic data could provide one approach to further estimating the utility of LLM-simulated expert data \citep{selby2025had,wilde2021_bayesiansynthetic}. 

Finally, the adoption of LLM-simulated experts introduces significant concerns regarding epistemic circularity, given that the instruments of estimation are fundamentally derived from the same architectures under assessment. As these systems achieve higher capability levels and potentially greater contextual awareness of evaluation regimes, their outputs may become increasingly susceptible to systematic bias or strategic misalignment with the objectives of objective risk quantification. This challenge is especially acute in modeling environments that lack robust external grounding or calibration against sparse empirical datasets. While such limitations do not necessitate the categorical exclusion of LLM-elicited estimates — particularly in light of the scarcity of scalable human alternatives — they underscore a critical research need for rigorous validation, methodological pluralism, and the development of hybrid elicitation frameworks that integrate automated estimation with expert human oversight.

Beyond technical performance, there are also questions of credibility and legitimacy in the use of LLM-based estimators. While LLM-based estimators may, in some contexts, match or even exceed human performance in forecasting tasks, their outputs may be viewed with greater skepticism by policymakers and other stakeholders. Human expert judgment, despite its limitations, carries institutional legitimacy that automated systems may lack. This creates a potential gap between what is technically feasible and what is socially or politically acceptable. Addressing this gap may require hybrid approaches that combine LLM-based estimation with human oversight, as well as greater transparency and validation to build trust in model outputs.

%%%%%%%%%%%% subsection %%%%%%%%%%
\subsection{Validating and Updating Models}

The utility of risk models must be maintained over time as the threat landscape, model capabilities, and institutional governance mechanisms all evolve. Workshop participants identified validation and updating as among the most technically underdeveloped aspects of the field, in part because standard tools for model validation assume a wealth of ground-truth data that does not exist for catastrophic AI risks. Additionally, much of the institutional infrastructure needed to support the validation and updating of these models has yet to be established.

A key challenge for validation arises from the sensitivity of model outputs to uncertain inputs, particularly in data-sparse environments. When parameter estimates are based on limited or low-quality data, even small changes in assumptions can lead to large differences in outcomes. In some cases, sensitivity analyses may reveal that model outputs are highly unstable, with conclusions shifting significantly across plausible parameter ranges. This raises concerns about overconfidence in point estimates and highlights the importance of systematically characterizing uncertainty, including through sensitivity analysis, scenario variation, and explicit representation of epistemic uncertainty.

%%%%%%%% sub-subsection %%%%%%%%%
\subsubsection{Validation and Model Quality Metrics}

Retrospective validation — the process of testing a model's predictions against observed outcomes — is the standard method for establishing that a risk model is accurate. In mature risk domains, historical incident data provides the basis for this kind of validation. For example, actuarial models of flood risk can be checked against flood records. AI risk modeling, in contrast, lacks this validation mechanism. The catastrophic events the models are designed to anticipate have not occurred, and any near-miss or precursor events that might serve as proxies are not well documented. As discussed in Section 4.3, even the cybersecurity domain, one of the most empirically grounded of the risk categories currently under consideration as it comes to catastrophic AI risks, suffers from severe data quality problems that make retrospective validation difficult.

This creates a fundamental epistemological challenge. A risk model that has never been validated against outcomes cannot be shown to be valid, but waiting for the outcomes that would enable validation is precisely what the models are designed to help avoid. The field, therefore, faces a difficult question about which validation standards are both achievable and sufficient to justify the use of AI risk models. The aim of risk modeling and risk assessment overall is not to perfectly predict future outcomes, but to equip decision makers with enough information to make reasonable decisions under uncertainty. 

However, without consensus on validation standards and approaches, we may find ourselves combining a variety of approaches and disparate data sources to create risk models of dubious value. Workshop participants did not resolve this tension, but identified several partial approaches worth pursuing, including developing internal model quality metrics and partial model validation.

Internal model quality metrics were proposed as a proxy for external validity in the absence of ground truth. Sensitivity analysis and Shapley value stability analysis, in particular, were noted as promising methods. These methods can be used to partially assess consistency and performance. For example, if the final risk output appears insensitive to large variations in a given parameter, that parameter's inaccuracy may matter less. If the relative importance ordering of variables is stable across different elicitation inputs, the model's structure can be said to at least be internally coherent. These properties do not guarantee that the model accurately represents real-world risk, but they reduce the model's vulnerability to input error and provide a principled basis for communicating where uncertainty is concentrated.

Partial model validation is the process of reviewing specific components or changes within a risk model rather than a full-scope assessment of the entire framework. It is typically employed when full validation is infeasible due to model complexity, but could also be useful for AI risk models, where validation is challenging due to limited ground truth. We envision that partial model validation could leverage the results of randomized controlled trials testing particular risk scenarios. For example, uplift studies are already used by frontier model developers and evaluators to gauge the extent to which releasing a new model will change the threat landscape. For example, these studies measure the capability gains novice hackers receive when given unrestricted access to a model and asked to complete a challenge \citep{fmf2025_capabilityassessments}. Data from these studies can be compared with estimates from the risk model before being integrated into it. Designing empirical studies that can partially validate model assumptions, even without end-to-end validation of the full risk space, represents a concrete and underexplored research direction.

%%%%%%%% subsection %%%%%%%%%%%%%%%%
\subsubsection{Model Assumptions Tracking}
A more tractable validation problem is maintaining explicit model assumptions over time. Workshop participants emphasized that risk models are built on a large number of assumptions. These range from the frequency with which threat actors attempt specific scenarios, the level of expertise required at each step of an attack chain, the effectiveness of defensive countermeasures, and the degree to which AI capabilities translate into operational uplift. These assumptions are typically fed into the model during model construction, but may become invalid as the threat landscape evolves, AI capabilities advance, or defensive methods change.

A system for tracking the assumptions underlying each model parameter, with their associated evidence and confidence levels, would allow model maintainers to monitor which assumptions are most likely to have been invalidated when new information arrives. Surveillance data, threat intelligence reports, incident databases, and API usage logs that monitor how models are used by threat actors were all identified as relevant information streams for this purpose. The feasibility of this monitoring depends in part on the institutional arrangements discussed in Section 4.5: who has access to which information streams and who has the mandate and resources to act on them.

The question of how frequently models should be updated did not reach consensus in the workshop, but participants generally agreed that updates should be triggered by substantive changes in the underlying evidence, rather than by fixed calendar schedules alone. The release of a significantly more capable model, the documentation of a novel attack campaign, the results of a new capability evaluation, or the invalidation of a core model assumption would all plausibly constitute update triggers. A purely calendar-based update cycle risks both under-updating in periods of rapid change and wasting resources on re-validation when conditions have not materially shifted.

%%%%%%%%%%%% subsection %%%%%%%%%%%%%%%%
\subsection{Governance and Institutional Design}

The workshop design also reflected that AI risk modeling cannot be treated as a purely technical exercise. The choice of institutional actors, disclosure norms, and oversight structures will shape not only the quality of the models but also what kinds of models are possible, what evidence can be gathered, and how widely the resulting assessments can be trusted and used.

%%%%%%%%%%% sub-subsection %%%%%%%%%%%%
\subsubsection{Roles and Responsibilities}

Questions of who should be responsible for constructing, maintaining, updating, and validating risk models were among the most actively deliberated in the workshop’s governance discussions. Several distinct issues emerged: \\

\textit{Domain expertise must be paired with contextual understanding} \\
Domain expertise emerged as indispensable, especially for identifying uplift and impact. But this expertise must be interpreted in the context of the relevant event chain to understand how harms materialize in practice. Participants noted that robust modeling requires not just specialized domain knowledge but also a strong understanding of how AI systems alter operational dynamics. These two types of expertise (narrow technical expertise, and broader contextual understanding) do not always reside in the same place, which can have direct implications on how teams should be assembled. \\

\textit{Institutional models are useful, but imperfect} \\
Several institutional models were discussed as possible analogues. Participants pointed to arrangements such as insurers, Underwriters Laboratories, and NIST-like bodies as examples of organizations that combine technical expertise with standardized evaluation practices. Others emphasized the value of independent third-party evaluators or verification organizations, ideally drawn from a pool of organizations that can be consulted without creating conflicts of interest \citep{homewood2025third}. \\

\textit{The same organization should not both support and audit the same model} \\
An obvious caveat was that the same organization should not both provide services to model developers and later audit them, suggesting a need for structurally independent evaluation and verification structures, potentially drawing on established models such as Independent Verification and Validation (IV\&V), to ensure credibility and avoid conflicts of interest. \\

\textit{Model developers and external evaluators should play distinct roles} \\
The workshop also highlighted a division of responsibilities between model builders and model users. Some participants argued that model providers should bear the burden of generating at least the baseline risk models, while external evaluators should be able to validate, challenge, or extend those models using independent information. This division becomes especially important when models are used for governance purposes such as threshold setting, where the legitimacy of the output depends not just on technical soundness but also on whether the process is perceived as sufficiently independent and contestable. \\

\textit{Information sharing requires tiered disclosure systems} \\
Another important concern that surfaced concerned what information should be shared, with whom, and under what protections. Workshop participants expressed support for sharing higher-level insights from risk models more broadly, while reserving more detailed scenario information, parameters, or structural components for trusted technical bodies. This reflects a tension familiar from vulnerability disclosure: disclosures can improve preparedness and accountability, but overly detailed release may also create misuse risks.

Participants suggested that structural information about risk models may often be more useful than raw parameter values, since the latter may be more sensitive and less transferable across contexts. At the same time, some argued that public-facing disclosures should be selective not only to reduce misuse risk, but also to allow different entities to receive different levels of detail depending on their security clearance and/legal mandate, through tiered disclosure regimes that permit sharing of distinct threat models with different audiences. \\

\textit{Liability protections are needed to help incentivise disclosure} \\
The workshop also surfaced a more basic governance challenge: liability. If a developer, evaluator, or government body discloses a risk scenario before a mitigation is in place, what protections exist against legal or reputational consequences? Participants suggested that safe-harbor arrangements could help align incentives for reporting and disclosure, particularly if the goal is to surface risks early enough for action. In the absence of such protections, actors may have strong reasons to suppress information that would otherwise improve collective risk awareness. 

%%%%%%%%% sub-subsection %%%%%%%%%%
\subsubsection{Oversight and Coordination}

Workshop participants noted the appeal of an institutional structure analogous to the Intergovernmental Panel on Climate Change (IPCC)~\citep{ipcc2023_ar6}. IPCC uses a variety of independently developed risk models, with an overarching assessment body responsible for synthesizing across them and communicating a centralized view. They also noted the IPCC's well-documented limitations as a governance mechanism. IPCC's consensus-based structure has made it vulnerable to political pressure from governments with interests in conservative risk estimates, and its outputs have historically understated the pace of realized climate change relative to the upper range of model projections.

A parallel structure for AI risk would face similar pressures, given the strong commercial and geopolitical incentives of frontier AI developers to present their systems' risks conservatively. The question of how to design an overarching assessment body that is institutionally insulated from these pressures without sacrificing the legitimacy that comes from broad participation is at least as much a political design problem as a governance or technical one.

Participants also emphasized the importance of monitoring systems over time, especially for updating assumptions as model capabilities, usage patterns, and threat environments change. This suggests a role for institutions that can collect incident reports, API usage data, and other monitoring signals, then convert them into triggers for review or revision. In some cases, participants suggested that governments may need to play a coordinating role in surfacing risk scenarios, sharing incident information, and maintaining reporting infrastructures, particularly for systemic risks that are diffuse and hard to observe directly. 

Participants also expressed the need for an independent (non-government non-industry) trusted third party entity to facilitate information sharing. They cited NASA’s Aviation Safety Reporting System (ASRS) \citep{nasa_asrs} as an example of a model that was established as a response to hesitation from companies to share data directly with regulators. This further highlights the need for an AI-focused Information Sharing and Analysis Center (ISAC) to act as a neutral repository for risk intelligence, and support industry-led rapid information sharing \citep{mitch2025_governanceapproaches}.  

%%%%%%% sub-subsection %%%%%%%%%%%
\subsubsection{Institutional Design Tradeoffs}

A broader takeaway from the workshop was that governance design choices cannot be separated from methodological choices. The level of model detail, the degree of transparency, the frequency of updating, and the choice of evaluator all shape one another. For example, a system that aims to support pre-registered risk thresholds will likely require simpler, more repeatable models and clearer institutional procedures than one aimed primarily at exploratory risk analysis. 

Participants also noted a likely tradeoff between broad participation and operational effectiveness. A highly participatory process may increase legitimacy, but it can also make consensus harder to achieve and create opportunities for strategic pressure. By contrast, a smaller and more specialized oversight structure may be more operationally useful, but may struggle to secure trust or access to the information needed for robust assessment. This suggests that future institutional design work should not aim to identify a single best model of governance, but rather to clarify which arrangements are appropriate for which decision contexts.

%%%%%%%%%%%% section %%%%%%%%%
\section{Prioritized Research Areas and Recommendations}
\label{sec:Prioritized}

\subsection{Prioritized Research Areas}
As a summary, a short list of areas where technical and governance research is likely to have an immediate benefit on AI risk modeling (based on Section 4) includes:

\textbf{Model Architecture and Structure}
\begin{itemize}
    \item Standards for evaluation reporting.
    \item Benchmarks tailored to specific risk scenarios, as opposed to just risk domains.
\end{itemize}

\textbf{Model Scope and Granularity}
\begin{itemize}
    \item Methods for assessing coverage of the risk universe and for characterizing whether resulting estimates should be interpreted as lower-bound, central, or upper-bound.
    \item Criteria for selecting an appropriate level of granularity given data sparsity, including when increased model complexity reduces rather than improves empirical robustness.
    \item Aggregation methods for combining outputs across ensemble modeling approaches, including how to triangulate directional signals and communicate disagreement when models diverge.
    \item Frameworks for matching model fidelity requirements to intended use (e.g., threshold-setting and safety assurance vs. mitigation prioritization or exploratory analysis).
\end{itemize}

\textbf{Evidence Integration and Elicitation}
\begin{itemize}
    \item Use of superforecasters instead of/in addition to domain experts.
    \item Validity of LLM-simulated experts across different risk domains and scenarios.
    \begin{itemize}
        \item Extend or validate  \citet{capstick2024autoelicit} on high-priority AI risk domains.
        \item Comparison of different models (primary importance).
        \item Comparison of different prompting methods and tasks (secondary importance).
    \end{itemize}
\end{itemize}

\textbf{Validating and Updating Models}
\begin{itemize}
    \item Reviewing standards for model quality in other domains and creating a set of model quality metrics for AI risk models. 
    \item Designing experiments for partial model validation.
\end{itemize}
 
\textbf{Governance and Institutional Design}
\begin{itemize}
    \item Division of responsibility between model builders and independent evaluators, including structurally adversarial arrangements and randomized assignment mechanisms to mitigate conflicts of interest \citep{homewood2025third}.
    \item Tiered disclosure regimes that distinguish structural model information from parameter-level information and calibrate access to recipient mandate and clearance.
    \item Safe-harbor arrangements to align incentives for good-faith disclosure of risk scenarios prior to mitigation deployment.
    \item Design of an overarching assessment body able to synthesize across independently developed models while remaining insulated from commercial and political pressure, drawing on the IPCC analogue and its documented limitations.
    \item Information-sharing infrastructures for incident, capability, and usage data, including the case for an AI-focused ISAC and an ASRS-style trusted third-party reporting channel \citep{mitch2025_governanceapproaches}.
\end{itemize}

%%%%%%%%%%%%%%%%
\subsection{Recommendations}
\begin{itemize}
    \item Establish independent evaluation and verification capacity
    \item Incentivize structured disclosure of model assumptions and evidence
    \item Use tiered transparency rather than all-or-nothing disclosure
    \item Create safe harbor protections for good faith reporting
    \item Build shared reporting and incident infrastructures
    \item Support an ensemble approach to modeling and oversight
\end{itemize}

%%%%%%%%%%%%%%%%%%%

\section{Conclusion}
As frontier systems grow more capable and more tightly coupled to high-stakes social and technical systems, the central challenge is not simply to measure capability, but to build better risk assessment tools that connect these capabilities to real-world harm pathways. Although no off-the-shelf risk modeling technique can be readily adopted to operationalize these efforts, probabilistic risk assessments, Bayesian causal inference, and threshold-based governance, among others, contribute important foundations to AI risk modeling. 

The workshop surfaced a set of open questions that cut across model architecture, scope, evidence integration, validation, and governance. Insights from the participants make it evident that progress on AI risk modeling will depend on parallel advances in technical methods and institutional design. The most important near-term priority is not to settle every open question, but to create governance arrangements that make iterative model development, validation, and review possible in practice.  The task ahead is to make that ecosystem more rigorous, transparent, and operational, so that quantitative AI risk modeling can evolve from a set of promising proposals into a reliable basis for governance.

%%%%%%%%%%%%%%%%%%%%%%%%%%%%%%%%
\section*{Contributions Statement} 
\textbf{Krystal Jackson, Malcolm Murray, Deepika Raman:} Conceptualization; Writing – original draft, Sections 1–6. 
\textbf{Jakub Kryś:} Writing – original draft, Section 4.
\textbf{Sean P. Fillingham, Jack Kengott, Andrew Lohn, Nada Madkour, Henry Papadatos, James Sykes, Anna Katariina Wisakanto:} Writing – review \& editing.
% \textbf{Krystal Jackson:} Conceptualization; Writing – original draft, Sections 1–6. 
% \textbf{Malcolm Murray:} Conceptualization; Writing – original draft, Sections 1–6.
% \textbf{Deepika Raman:} Conceptualization; Writing – original draft, Sections 1–6.
% \textbf{Jakub Kryś:} Writing – original draft, Section 4.
% \textbf{James Sykes:} Writing – review \& editing.
% \textbf{Anna Katariina Wisakanto:} Writing – review \& editing.
% \textbf{Sean P. Fillingham:} Writing – review \& editing.
% \textbf{Andrew Lohn:} Writing – review \& editing.
% \textbf{Nada Madkour:} Writing – review \& editing.
% \textbf{Henry Papadatos:} Writing – review \& editing. \\
% 
All authors reviewed and approved the final manuscript.

\section*{Acknowledgments}
The authors thank Jane E. Valentine, Kamile Lukosiute, Colin Shea-Blymyer, Richard Mallah, and Giovanna Jaramillo-Gutierrez for reviewing the manuscript.

%%%%%%%%%%%%%%%%%%%%%%%%%%%
%\section*{References}
\bibliographystyle{unsrtnat}
{\small
\bibliography{references}
}

\end{document}